\documentclass[useAMS,usenatbib, usegraphicx, epsfig]{mn2e}

\usepackage{amssymb}
\usepackage{subfigure}
\usepackage{amsmath}
\usepackage{multirow}
\usepackage{url}
\usepackage{color}

\title[Mitigating Angular Systematics]{Mitigating Systematic Errors in
  Angular Correlation Function Measurements from Wide Field
  Surveys\thanks{Based on observations obtained with
    MegaPrime/MEGACAM, a joint project of CFHT and CEA/DAPNIA, at the
    Canada-France-Hawaii Telescope (CFHT) which is operated by the
    National Research Council (NRC) of Canada, the Institut National
    des Sciences de l'Univers of the Centre National de la Recherche
    Scientifique (CNRS) of France, and the University of Hawaii. This
    work is based in part on data products produced at TERAPIX and the
    Canadian Astronomy Data Centre as part of the Canada-France-Hawaii
    Telescope Legacy Survey, a collaborative project of NRC and
    CNRS.}}

\author[C.~B.~Morrison et al.]{C.~B.~Morrison$^{1}$\thanks{E-mail: cmorrison@astro.uni-bonn.de},
H.~Hildebrandt$^{1}$ \\ 
$^{1}$Argelander-Institut f\"ur Astronomie, Auf dem H\"ugel 71, 53121 Bonn, Germany\\
}

\begin{document}

\maketitle

\begin{abstract}

  We present an investigation into the effects of survey systematics
  such as varying depth, point spread function (PSF) size, and
  extinction on the galaxy selection and correlation in photometric,
  multi-epoch, wide area surveys. We take the Canada-France-Hawaii
  Telescope Lensing Survey (CFHTLenS) as an example. Variations in
  galaxy selection due to systematics are found to cause density
  fluctuations of up to 10\% for some small fraction of the area for
  most galaxy redshift slices and as much as 50\% for some extreme
  cases of faint high-redshift samples. {This results in
  correlations of galaxies against survey systematics of order $\sim$1\% when 
  averaged over the survey area.}
  We present an empirical method for mitigating these
  systematic correlations from measurements of angular correlation
  functions using weighted random points. These weighted random
  catalogues are estimated from the observed galaxy over densities by
  mapping these to survey parameters. We are able to model and
  mitigate the effect of systematic correlations allowing for
  non-linear dependencies of density on systematics. Applied to
  CFHTLenS we find that the method reduces spurious correlations in
  the data by a factor two for most galaxy samples and as much as an
  order of magnitude in others. Such a treatment is particularly
  important for an unbiased estimation of very small correlation
  signals, as e.g. from weak gravitational lensing magnification
  bias. {We impose a criterion for using a galaxy sample in a
  magnification measurement of the majority of the systematic
  correlations show improvement and are less than 10\% of the expected
  magnification signal when combined in the galaxy cross correlation. After
  correction the galaxy samples in CFHTLenS satisfy this criterion for $z_{\rm
  phot}<0.9$ and will be used in a future  analysis of magnification.}

\end{abstract}



\section{Introduction}

Measurements of the angular correlation of galaxies are powerful tools
for probing the properties of galaxies and cosmology. 2-point
correlation functions such as galaxy auto-correlation functions, weak
gravitational lensing shear and weak lensing magnification are the
basic observables of many current and future photometric
surveys. However, the signals in these correlations can be
contaminated by systematic errors arising from spatially inhomogeneous
selections. Measurements of magnification bias especially suffer from
these effects due to its small amplitude, even at small scales. While
this paper takes systematics in magnification bias measurements as an
example, the methods described herein will be useful for any small- to
intermediate-scale, configuration-space correlation measurement.

Magnification has shown utility in measuring both mass and
cosmology. Many of these measurements have been of magnification bias,
the density change of galaxies due to lensing by intervening mass
\citep{ford12, ford14, hildebrandt09b, hildebrandt11, hildebrandt13,
  morrison12, scranton05, bauer11}. Other measurements have been also
performed using magnitudes and sizes \citep{bauer11, bauer14, huff11,
menard10, schmidt12}. These measurements have been shown to be complementary
to other probes of lensing of large-scale structure
\citep{vanwaerbeke10, menard02}. Much of this complementarity is from
the magnification signal coming for free, that is the same survey
design and data reduction for weak lensing shear surveys also allows
for magnification. Indeed, weak lensing magnification does not rely on
shape measurements meaning that it will not suffer from the same
systematic errors. This comes at the cost of greater sensitivity to
photometry errors and selection effects induced by survey systematics
such as depth and PSF size. While magnification shows large
sensitivity to these selection effects they will also be present in
other correlation analyses including weak lensing shear though as
second-order effects.

Much of magnification's and other correlations' mass and cosmological
constraining power comes from measuring the signal at different radial
distances and hence cosmic time. Selecting galaxies at a given
distance is commonly done through colour and magnitude cuts or more
generally photometric redshifts (photo-$z$). These selections should
be uniform across the survey to maximize their scientific
utility. However, variations in selection efficiency across the survey
can induce spurious systematic errors into the correlations. This
selection function can be caused by variations in survey depth, PSF
full width half max (FWHM), extinction, and stellar density, for
example. The different colour and photo-$z$ selections cause the
inhomogeneities to be different for each galaxy sample considered.

Current and future multi-epoch surveys will all suffer from such
varying selection functions that will contaminate correlation
measures. By leveraging the statistical power of the
Canada-France-Hawaii-Telescope Lensing Survey (CFHTLenS) dataset, we
investigate the effect of depth, PSF size, and extinction on galaxy
density for several different galaxy samples. We find that these
systematics can have an influence the density, changing it by 10\% in
some small fraction of the survey area for most samples and up to 50\%
for some faint high-$z$ samples. We then build an empirical,
cosmology-independent method for reducing the effect of these
systematics on correlation measurements. Techniques developed on
CFHTLenS can be immediately used on other similar surveys such as the
Red Sequence Cluster Lensing Survey
(RCSLenS)\footnote{http://www.rcslens.org} and the Kilo Degree Survey
(KiDS)\footnote{http://kids.strw.leidenuniv.nl/} as well as any survey
using a data pipeline and survey strategy similar to CFHTLenS.

The paper is organized in the following way. The CFHTLenS data are
presented in Sect.~\ref{sec:data}. Sect.~\ref{sec:galaxy_density}
discusses angular systematics and how we relate those to the galaxy
density. Results are presented in Sect.~\ref{sec:results} where we
show the galaxy densities as a function of the different systematics
and the resulting spurious correlations. In section
Sect.~\ref{sec:remove_sys} we develop our method for modeling and
removing angular systematics with weighted random points. In
Sect.~\ref{sec:results_cor}, we show the results from applying this
correction to the CFHTLenS dataset. Finally, in
Sect.~\ref{sec:conclusions} we summarize and give an outlook to
applications within CFHTLenS and beyond.


\section{Data}\label{sec:data}

\begin{figure*}
    \includegraphics[width=0.497\textwidth]{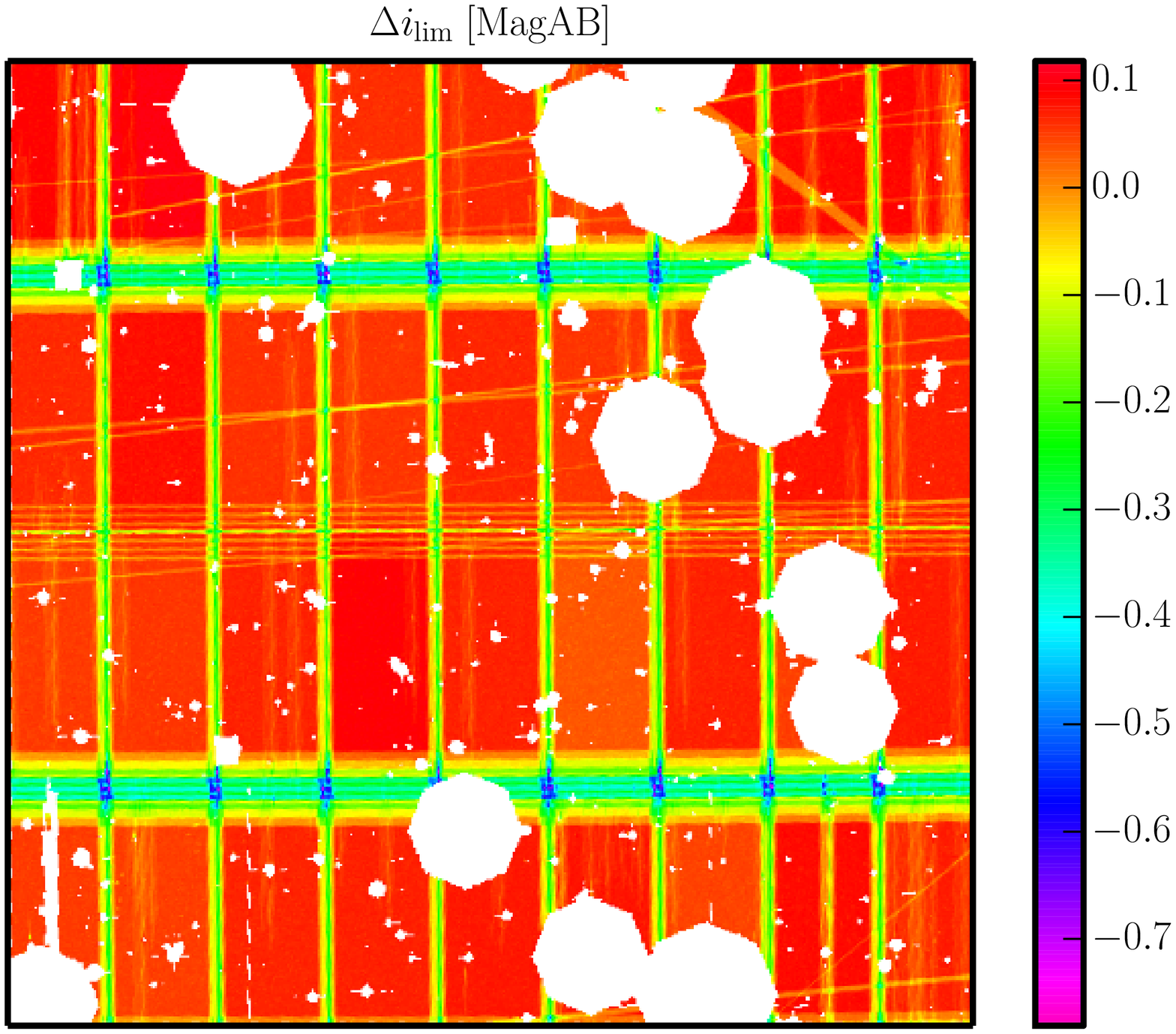}
    \includegraphics[width=0.497\textwidth]{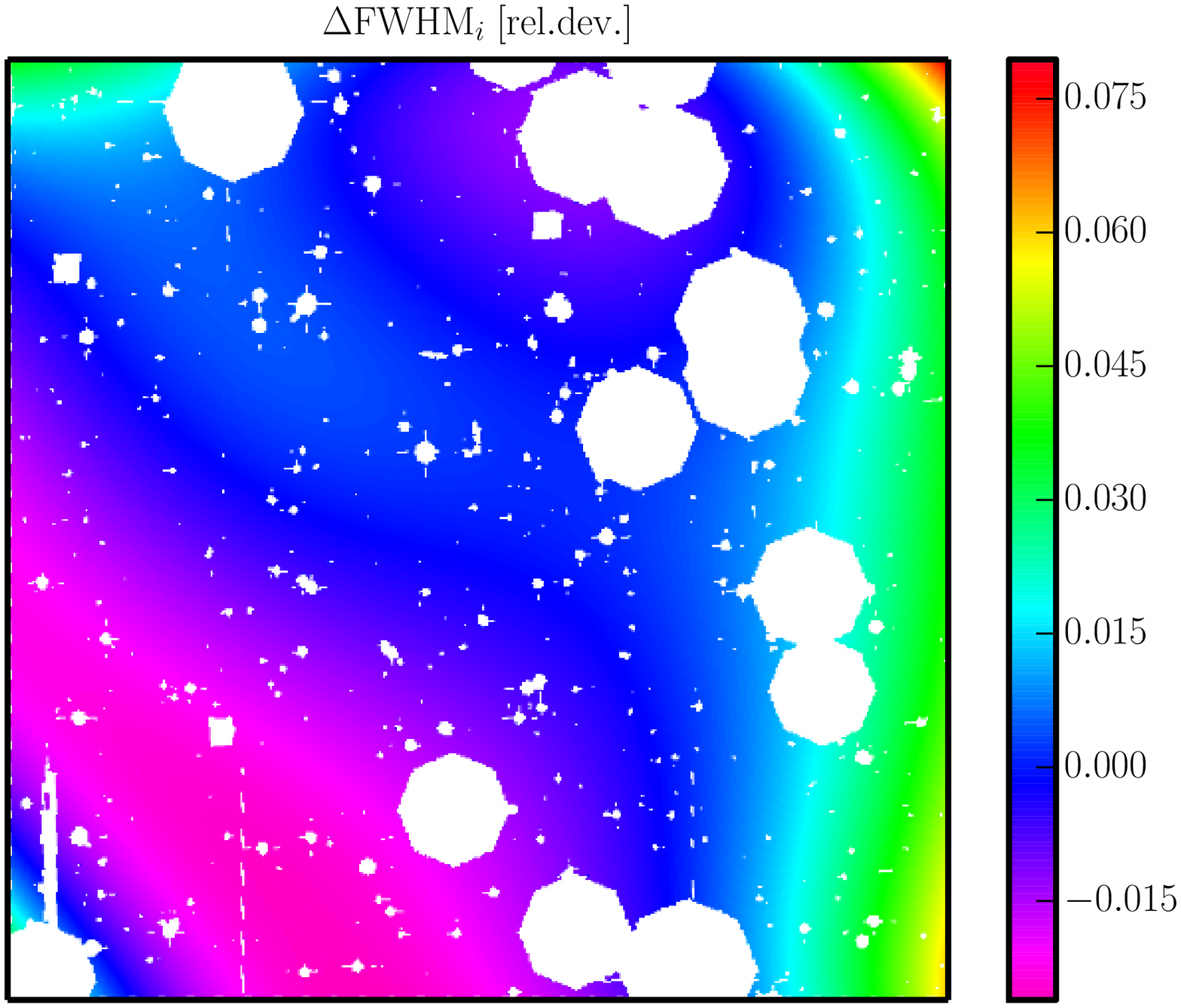}
    \caption{\label{fig:sys_maps_ds}Map of the $i$-band depth (left)
      and PSF size (right) for one CFHTLenS pointing
      ($\sim1$ sq. deg.). Each side of this plot is 1 sq. deg.
      {Values for the depth are relative to the mean of the
      field with the PSF size being fraction minus 1 relative to the mean.} The
      depth shown is estimated from the inverse variance image produced
      during stacking. Red corresponds to deeper regions whereas green and
      blue correspond to shallower regions. The few blue pixels roughly one
      magnitude shallower than the red regions. The PSF size is estimated
      from the model used for the PSF homogenization of \citet{hildebrandt12}.
      The colour scale shows large PSF in red/green and small PSF in
      blue/purple.}
\end{figure*}

The CFHTLenS data are described in \cite{heymans12}, \cite{erben13},
\cite{miller13} and \cite{hildebrandt12}. Here we will concentrate on
the aspects of the data used in this analysis. CFHTLenS is a 5 band
($ugriz$) imaging survey consisting of 171 pointings of
$\sim1$ sq. deg. each observed in 4 different regions on the sky.
Within each pointings the data are observed using 4-7 stepped dithers depending
on the band pass. This observing strategy yields little overlap between
pointings and CCDs. The unmasked area of the survey covers 117 sq. deg. This
area represents the strictest masking available for the survey. Reduced images
and catalogues were produced using the THELI-pipeline \citep{erben05}. Objects
are detected in the $i$-band and the multi-colour photometry is corrected for
different PSF sizes between bands by performing a PSF homogenization
\citep{hildebrandt12}.

Galaxies in this analysis are selected in two broad samples. The
first, hereafter referred to as photo-$z$ samples, are selected to have
$19 < i < 24$ magnitudes with photometric redshifts selected from
$BPZ$ \citep{benitez00} best fit redshift, $z_b$. The other samples,
hereafter referred to as LBGs, are selected using the dropout criteria
of \citet{hildebrandt09a} and have magnitude cuts of $23 < r < 24$ and
$23.5 < i < 24.5$ for the $u$- and $g$-dropouts, respectively. These
samples range from a total of $\sim$1 million galaxies in some of the
photo-$z$ samples to $\sim$30 thousand galaxies in the LBG samples.

For this analysis we consider 11 different survey systematics in
total. We consider the limiting magnitude, also referred to as the
depth, in each of the five observed bands in the survey. The depths
are estimated from the inverse variance images created during the
THELI stacking process. We also use the PSF size in each of the five
bands. We utilise the sizes as modeled for the PSF homogenization of
\citet{hildebrandt12}. The stellar density in the survey determines
the fidelity of the PSF model. Small-scale variations can not be
captured due to the limited amount of information and the smoothing
due to the functional fit. However, we do not expect the size of the
stacked PSF to vary too much at small scales. As the last survey
systematic we consider galactic extinction which we estimate from the
$E(B-V)$ values from the Schlegel-Finkbeiner-Davis
\citep[][SFD]{schlegel98} dust map.

To encode the survey geometry and and position dependence of the
systematics we utilise the spherical pixelisation code STOMP
\citep{scranton02}\footnote{http://code.google.com/p/astro-stomp}. This
software allows us to pixelate the area of the survey and include
complicated masking. We create constant resolution STOMP maps encoding
the number of galaxies as a function of position for a galaxy sample.
{The number counts of galaxies are sample at pixels of 16 sq.
arcminutes and are summed or smoothed as need requires.}.
We also create constant resolution maps that encode variations in
depth, PSF FWHM, and extinction. In Fig.~\ref{fig:sys_maps_ds} we show
the $i$-band depth and seeing for a $\sim1$ sq. deg. pointing in
CFHTLenS. One can clearly see the effect of the dither pattern in the
depth map. This pattern is repeated in the other CFHTLenS bands which
leads to a strong correlation of the depth in the different
bands. Note that the shallowest pixels are up to one magnitude
shallower than the mean of the pointing. There are also smaller scale
structures from pixel to pixel sensitivity differences. These
variations change the number density and properties of the detected
galaxies in the $i$-band and also affect the colours of galaxies
through the depth of the other four bands. The PSF size varies at much
larger scales as shown in the right hand panel of
Fig.~\ref{fig:sys_maps_ds}. This affects both the detection efficiency
(in the $i$-band) as well as the colour selection (in all bands)
because the seeing influences the noise for small objects. Larger PSFs
mean the detected surface brightness of galaxies decreases which
affects errors and possibly also magnitude estimates through noise
bias. {Some of these colour differences are corrected in the PSF
homogenization of \citep{hildebrandt12}, however, the noise variations
associated with the varying PSF will remain and affect selections.}

\begin{figure}
    \includegraphics[width=0.5\textwidth]{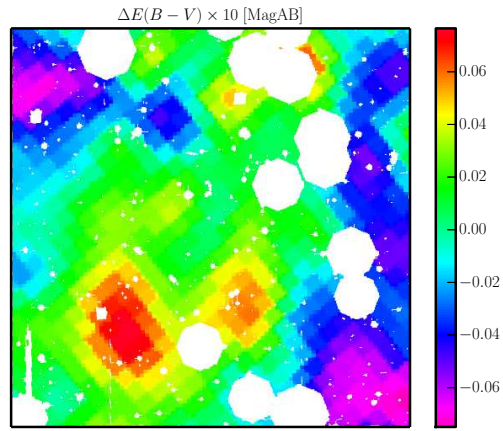}
    \caption{\label{fig:sys_maps_ext}Map of $E(B-V)$ extinction
      relative to the mean of the pointing for the same CFHTLenS pointing as
      shown in Fig.~\ref{fig:sys_maps_ds}. Red represents regions of high
      extinction blue and purple regions of low extinction compared to the
      pointing mean. The values for the extinction are taken from the galactic
      dust maps of \citet{schlegel98} over-sampled to the same resolution as the
      depth and seeing maps.}
\end{figure}

Fig.~\ref{fig:sys_maps_ext} shows the extinction for the same
pointing seen in Fig.~\ref{fig:sys_maps_ds}. The galactic dust causes
a dimming and a reddening of galaxy colours. While CFHTLenS corrects
for the reddening effect the dimming leads to a position dependent
depth variation with the same consequences as the depth variation due
to the dithering discussed above. The commonly used dust maps are also
contaminated by light from high-redshift galaxies. This can lead to an
unwanted physical correlation between the extinction correction and
cosmic density.

We also encode the density of stars from two different star
selections from \citet{hildebrandt12} and \citet{miller13}. In the
end, however, we do not utilize the stellar densities in this analysis
due to difficulties with star-galaxy separation as described in
Sect.~\ref{sec:stellar_den}. 

\section{Galaxy Density Estimates}\label{sec:galaxy_density}

Angular systematics in clustering measurements can arise from a
position-dependent selection function caused by variations in
instrumental and astrophysical quantities. The position-dependence
gets imprinted on the data. Mathematically this can be written as
\begin{equation}
 \delta{(\vec{\theta})} = \frac{S(\vec{\theta})N(\vec{\theta}) - 
 \langle S(\vec{\theta})N(\vec{\theta}) \rangle}
 {\langle S(\vec{\theta})N(\vec{\theta}) \rangle}
\end{equation}
where $\delta(\theta)$ is the galaxy over-density at position $
\vec{\theta}$, $N$ is the true galaxy number count, and $S$ is the
sample-dependent selection. This factor can be significant, changing
the density by as much as $\sim 50\%$ for certain samples in
CFHTLenS. In the remainder of the paper we distinguish between
$\delta$ and $\hat\delta$ where the latter quantity is our estimate
from the CFHTLenS data, i.e. including shot-noise from the finite
{density} of the survey. CFHTLenS is split up into individual
pointings of roughly 1 sq. deg. each. {Within these pointings the
data are observed using 4-7 stepped dithers depending on the band pass. This
observing strategy yields little overlap between pointings and CCDs as can be
see in Fig.~\ref{fig:sys_maps_ds}. In this figure, green and blue regions are
chip gaps between CCDs which are significantly shallower than the mean of the
survey by 0.2 magnitudes. Further information on the dither pattern can be found
in \citet{erben13}}. This limited overlap is a feature shared
with many other current and near future surveys. This can lead to
difficulties in the photometric calibration between pointings,
inducing changes in galaxy densities related only to calibration
issues. Analyses interested in scales larger than the pointing size
need to be corrected for this effect. We instead remove variations
that occur from pointing to pointing as we are interested in scales of
the size of roughly a galaxy halo, much smaller than the CFHTLenS
pointing size. Practically, we utilize the galaxy over-density
relative to the mean of the pointing to compute our statistics instead
of the global survey mean. We similarly do this with each systematic
considered, i.e. we relate the depth/seeing/extinction in a particular
pixel to the mean of the depth/seeing/extinction in the pointing that
the pixel resides in.

\begin{figure}
    \includegraphics[width=0.495\textwidth]{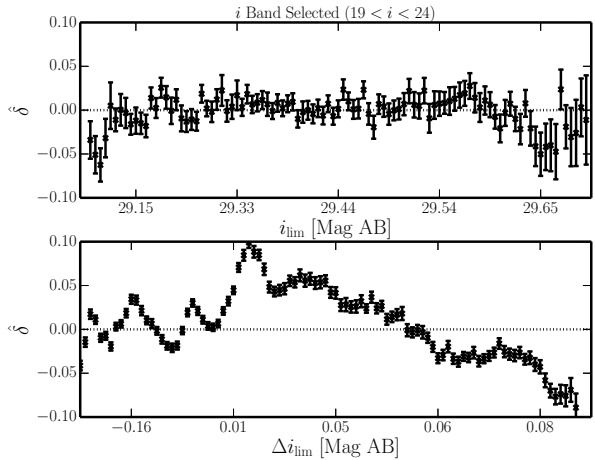}
    \caption{\label{fig:den_compare} Galaxy over-density $\hat\delta$
      as a function of $i$-band depth for an $i$-band selected sample
      with $19 \le i < 24$. {The top panel shows the over-density
      with respect to the survey mean galaxy density against the limiting
      magnitude (per pixel). The bottom panel shows the same but using
      galaxy densities values relative to the mean in the individual pointings,
      averaged over the pointings as described in \S~\ref{sec:galaxy_density}.
      Each data point corresponds to a non-contiguous survey area of 1~sq.~deg,
      that is from all pointings in the survey containing the
      systematic value range. The x-axis is plotted on a non-linear scale to
      allow for equal spacing between the data-points to show the variations in
      density more clearly around the mean value of depth.}}
\end{figure}

In Fig.~\ref{fig:den_compare} we compare the dependence of the
measured over-density on $i$-band depth before and after going from
absolute (i.e. relative to the survey mean) to relative (i.e. relative
to the mean of individual pointings) values. {Both the galaxy density
and $i$-band depth are estimated on a scale of $8$ arc-seconds (we use this
smoothing for all subsequent density estimates).} Note the large increase
in signal to noise between the two panels. While we show this only for
the $i$-band depth this is also true for the other systematic
quantities. This is probably due to the data reduction being done
pointing-wise and the catalogues being created from running
SExtractor\citep{sextractor} on the individual pointings rather than
running the detection over the full survey at once. We will utilize
this relative scaling for all plots in the remainder of the paper.
 
\section{Dependence of galaxy density on
  systematics}\label{sec:results}

First we compare the over-density of galaxies in each sample against
each of the systematics we consider separately. We bin the relative
value of the systematic in ranges that represent equal area on the
sky. We compute the values and errorbars by spatially jackknifing over
the pointings of CFHTLenS (171 in total) contributing to each bin. This
jackknifing is done with weighting by area, {that is pointings
treated equally in the jackknife without regard for how much area they
contribute to a particular systematic bin}. The plots in the following sections
show how the density can depend on the different systematics even for relatively
bright, low redshift galaxy samples.


\subsection{$i$-band Selected Sample}\label{sec:i_band_selected}

\begin{figure}
    \includegraphics[width=0.53\textwidth]{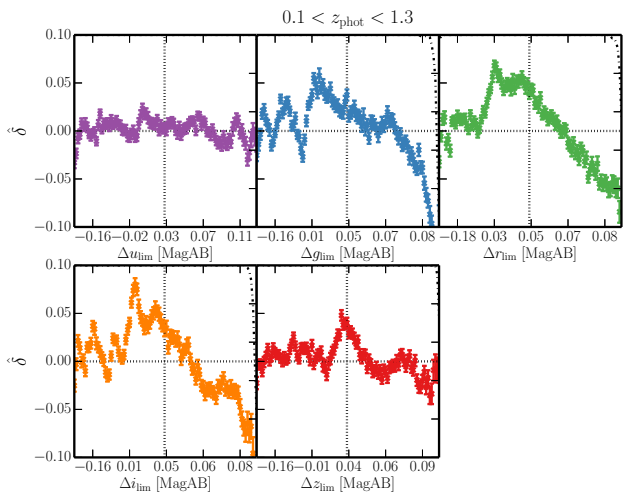}
    \caption{\label{fig:z01t13_sky_den}Plots of the galaxy
      over-density $\hat{\delta}$ against the relative depth in the
      five CFHTLenS optical bands for the sample $0.1<z_{\rm phot}<1.3$. The
      dot dashed line visible for large values of depth is the percentage of
      pointings (ranging from all 171 to 0) contributing to the density
      estimate of each depth bin.}
\end{figure}

\begin{figure*}
    \includegraphics[width=0.95\textwidth]{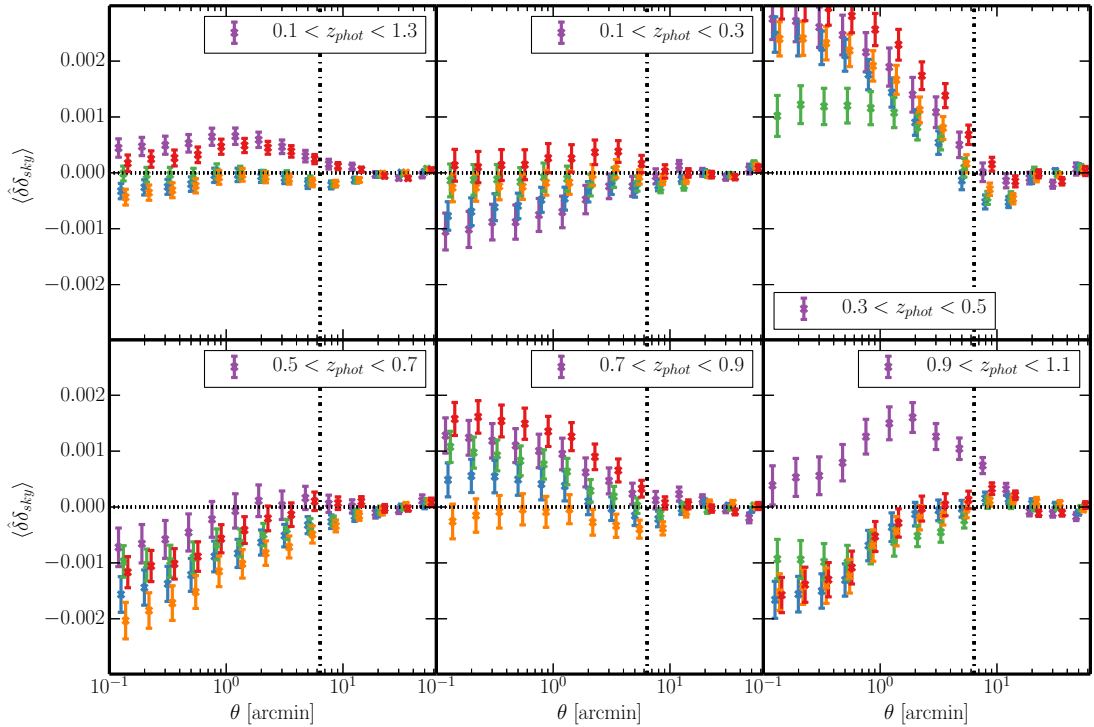}
    \caption{\label{fig:fore_sky}Angular cross correlation functions
      between the observed galaxy density $\hat{\delta}$ and the
      survey depth $\delta_{sky}$ in each band {where purple$=u$,
      blue=$g$, green=$r$, yellow=$i$, red=$z$.} We keep this
      convention throughout the paper.  We plot all photo-$z$ selected
      samples up to $0.9<z_{\rm phot}<1.1$ including the total sample
      of $0.1<z_{\rm phot}<1.3$ (top left). We neglect the $1.1<z_{\rm
      phot}<1.3$ for brevity though it is similar to the $0.9<z_{\rm phot}<1.1$
      sample. The vertical dot-dashed line corresponds to the width of
      a MegaCam Chip.}
\end{figure*}

Before presenting results using photo-$z$ and colour-colour
selections, we show the density variations of a purely $i$-band
selected sample plotted against the $i$-band depth. The lower panel of
Fig.~\ref{fig:den_compare} shows that the density as a function of
relative $i$-band depth fluctuates in a complex way. For the deeper
part data (right half of the lower panel of
Fig.~\ref{fig:den_compare}), we can understand this as the effect of
Eddington bias leading to a monotonic decrease in density with
increasing depth. {This begins around the value of $\Delta i >
0.05$.} In general, the Eddington bias in this case
describes the asymmetric scatter of objects close to the noise
limit. Since there are more faint than bright objects more objects
scatter into the sample than out of the sample at the faint end. As
the depth increases this effect becomes less important and density
decreases. This is a general trend seen in the other bands as well.
{We have no such explanation for the shallower portions of the
survey} (left half of the lower panel of Fig.~\ref{fig:den_compare})
though the trends seen there are likely due to complexities in the
stacking and detection pipeline. {While it is tempting to claim that
the $i$-band selected sample is completely independent of effects from the other
five bands due to it's selection relying on $i$-band only, this is not the case
here as the survey mask used in this analysis uses information from all the
observed bands and could introduce a selection dependent on the other
bands. This can be seen in the next section where we plot these density
fluctuations for the sample $0.1<z_{\rm phot}<1.3$ which is very similar to the
$i$-band selected sample.}


\subsection{Galaxy Density vs. Depth}\label{sec:den_vs_depth}

We now show the plots of galaxy density against the depth in all five
bands for samples of galaxies in CFHTLenS selected in photometric
redshift bins. In Fig.~\ref{fig:z01t13_sky_den} we show the density
variations of the $0.1<z_{\rm phot}<1.3$ sample against the depth. The
dependence on $i$-band depth here is very similar to that of
Fig.~\ref{fig:den_compare} showing an amplitude of up to
$\sim$10\%. The other bands also show significant structure, however,
some of this is expected to be redundant as the dither patterns are
similar and therefore the depth varies in a similar way between the
bands. Indeed upon cross-correlating the depth maps of different bands
with each other we find that the amplitude is roughly 50\% of the
amplitude of auto-correlation of the individual band depth maps. For
the low and high redshift galaxy samples that are a subsets of the
sample shown in Fig.~\ref{fig:z01t13_sky_den}, the density variations
in the $u$ and $z$-bands become more apparent as these bands are
important for distinguishing high and low redshifts.

The cross-correlations of the galaxy positions for this sample against
the depth maps are shown in the top-left panel of
Fig.~\ref{fig:fore_sky}. Correlations are small here ($<0.1\%$) since
no aggressive colour cuts are applied. The remaining panels of
Fig.~\ref{fig:fore_sky} show the same correlations for narrow
photo-$z$ slices. {Much larger amplitudes of up to $\sim0.3\%$ occur
for some samples indicating the importance of the colour selection. Most of the
correlations have an amplitude of $\sim0.1\%$ though and do not 
show significant structure as a function of $\theta$. For the depth, we have
information down to the scale of roughly an arc-second and the
auto-correlations of the depth maps do show power at these small sales. The
galaxy samples show sensitivity to these small scales as some of the
correlations (such as the $0.5<i_{\rm phot}<0.7$ sample) are not leveling off
as they would if there was no dependence.}

Two samples of note are the $0.3<z_{\rm phot}<0.5$ sample and the high
redshift $0.9<z_{\rm phot}<1.1$ sample. The $0.3<z_{\rm phot}<0.5$
sample shows significantly higher cross-correlations against the depth
maps - by about a factor of two - compared to the other samples. One
would expect that this sample would be mostly free from systematics
due to its low redshift and bright nature (i.e. similar to the
$0.1<z<0.3$ sample). This is not the case as selecting this sample by
photo-$z$ relies heavily on information from the $u$-band, which is
shallower compared to the detection band, to distinguish it from the
higher redshift selections. This is also true for higher redshifts
such as $0.9<z_{\rm phot}<1.1$, which use a lot of information from
the $u$- and $z$-bands for the redshift determination. Coupled with
the faintness of the sample the $0.9<z_{\rm phot}<1.1$ sample exhibits
a unique behavior in that the cross-correlation against the $u$-band
depth map is significantly different from the cross-correlations
against the other bands, with an opposite sign and a large amplitude.

\begin{figure}
    \includegraphics[width=0.495\textwidth]{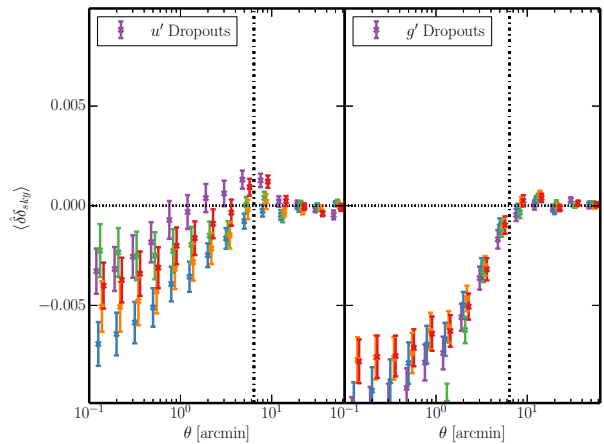}
    \caption{\label{fig:back_sky}Similar to Fig.~\ref{fig:fore_sky}
      but for samples of Lyman-break Galaxies (LBGs). Note the
      different y-axis range.}
\end{figure}

Fig.~\ref{fig:back_sky} shows the angular cross-correlation function
between the density of the two LBG samples and the depth in the
different bands. These galaxy samples show significantly larger
density variations (not shown), up to five times that of the brighter
photo-$z$ selected samples. Likewise we observe a roughly 10 fold
increase in the amplitude of the cross-correlations for these
samples. This drastic increase is to be expected given the faint
nature of the LBG samples and their sensitivity to changes in the
depth. The correlations that result from this are largely monotonic in
$\theta$ with significant correlation at small and intermediate
scales. These large correlations, in combination with the correlations
from the other foreground samples, could lead to significant
contamination of cross-correlation measurements between the different
galaxy samples such as the ones used for magnification bias.

\subsection{Galaxy Density vs. PSF Size}

\begin{figure}
    \includegraphics[width=0.53\textwidth]{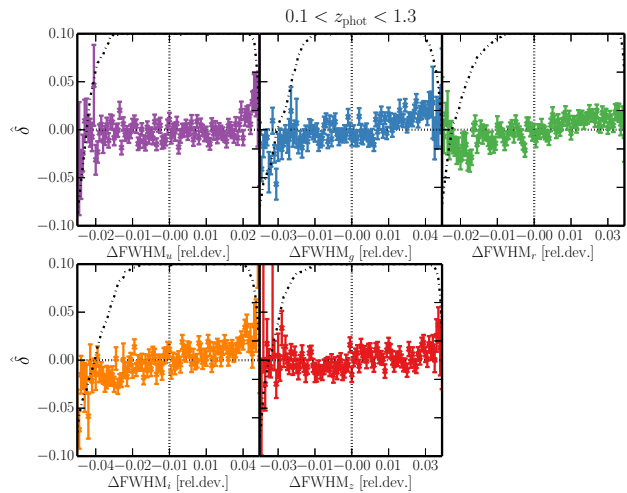}
    \caption{\label{fig:z01t13_see_den}Plots of the galaxy
      over-density $\hat\delta$ as a function of the PSF FWHM in the
      CFHTLenS optical bands for the sample $0.1<z_{\rm phot}<1.3$.}
\end{figure}

\begin{figure*}
    \includegraphics[width=0.95\textwidth]{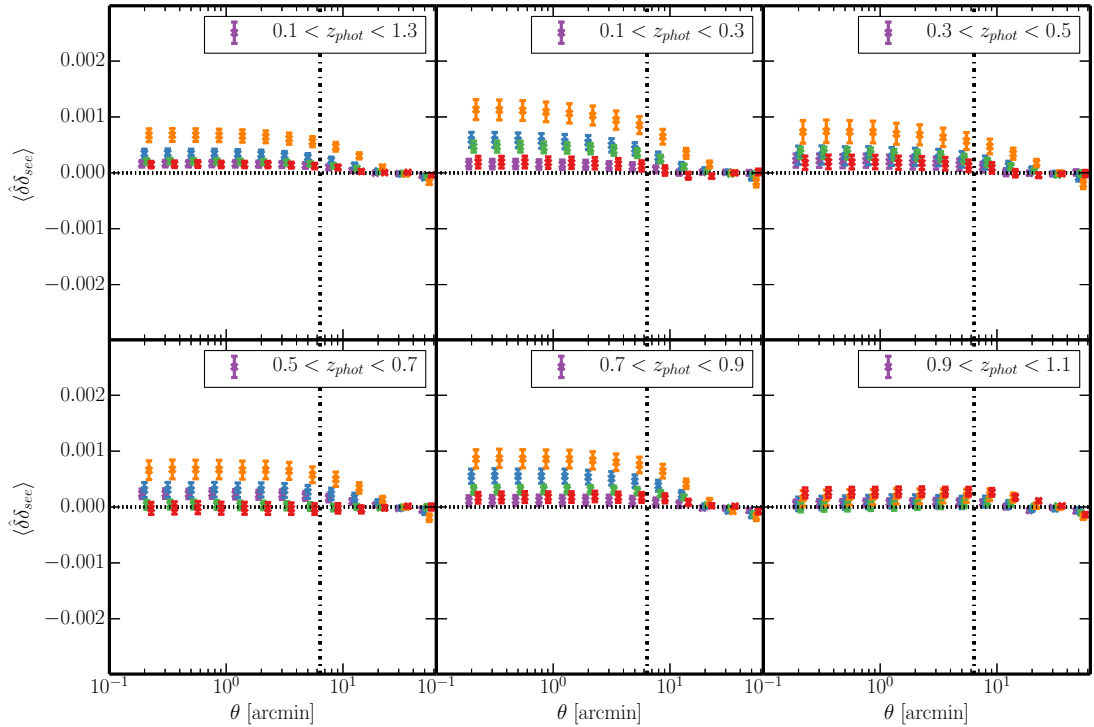}
    \caption{\label{fig:fore_see}This figure is similar to
      Fig.~\ref{fig:fore_sky} except we now show the correlations
      against the PSF FWHM in each of the five bands.}
\end{figure*}

In this section we analyze  the dependence of the galaxy density on the
PSF size. As shown in Fig.~\ref{fig:z01t13_see_den} these density
fluctuations are less pronounced compared to the trends seen with
survey depth (Sect.~\ref{sec:den_vs_depth}). They remain significant
however with fluctuations of around 5\% at the highest. The strongest trend in
the density is found when looking at the $i$-band PSF size. This is likely
due to the fact that the $i$-band is the detection band. This also
suggests that the PSF homogenization that corrects for colour effects
from different PSF sizes in each band \citep{hildebrandt12} is working
properly.

Correspondingly Fig.~\ref{fig:fore_see} shows the cross-correlation of
the galaxy densities in the different samples against the PSF FWHM
maps. {The majority of the correlation seen here is flat at smaller
scales ($\theta < 1 {\rm arcmin}$). This is not unexpected due to the
fidelity of the seeing model we use from \citet{hildebrandt12}. We do not expect
the PSF size to be much of an issue at the smallest scales as it tends to vary
much more smoothly compared to the depth. The amplitude we observe is as large
as that of the depth} and the correlation amplitude is almost always largest
against the $i$-band PSF size map indicating that much of the correlation from
the PSF is simply due to the detection process alone. Interestingly,
the highest redshift slice appears to perform contrary to this trend
with the correlation against the $z$-band PSF FWHM map being
largest. This again suggests that the higher redshift samples are more
sensitive to the $z$-band through the redshift selection.

\begin{figure}
    \includegraphics[width=0.495\textwidth]{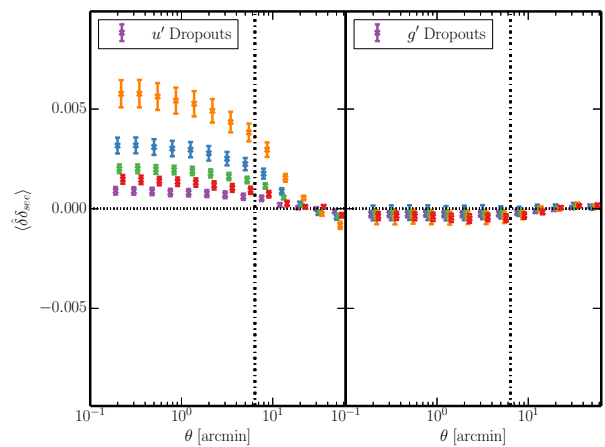}
    \caption{\label{fig:back_see}This figure is similar to
      Fig.~\ref{fig:fore_see} except we now show the
      cross-correlations of the densities of the two LBG samples
      against the PSF FWHM maps in the different bands.}
\end{figure}

The results are mixed for the LBGs as shown in
Fig.~\ref{fig:back_see}.  The $u$-dropouts show large correlation
amplitudes roughly equal in amplitude to the correlations against the
depth maps. However, the $g$-dropouts, which have the largest
cross-correlation amplitudes to the depth overall, show very little
correlation to the PSF size, with the amplitude being an order of
magnitude lower that that of the $u$ dropouts. We do not have an
explanation for this surprising behaviour.

\subsection{Galaxy Density vs. Extinction}

Milky-Way dust can also introduce spurious correlations into galaxy
samples. High redshift galaxies observed through galactic dust will
have their colours reddened and thus their selection will change as
the dust opacity changes. This effect usually is corrected using the
Schlegel-Finkbeiner-Davis \citep[SFD][]{schlegel98} dust maps or
similar data, modifying the galaxy sample's colours based on the
amount of dust measured at a given location. However, these
corrections cannot account for changes in the noise properties of the
detected galaxies or - equally important - any galaxies not detected
due to the extra opacity. SFD and similar maps of dust opacity are
also contaminated by high redshift galaxies as the infrared emission
used to determine the dust opacity is contaminated by the emission
from the Cosmic Infrared Background \citep[CIB,][]{schmidt15}. In
addition to this, we do not have access to data describing the small
scale fluctuations of dust extinction. Indeed, the small scale
structure of galactic extinction is an ongoing topic of research
\citep{planck13xi}.

\begin{figure}
    \includegraphics[width=0.495\textwidth]{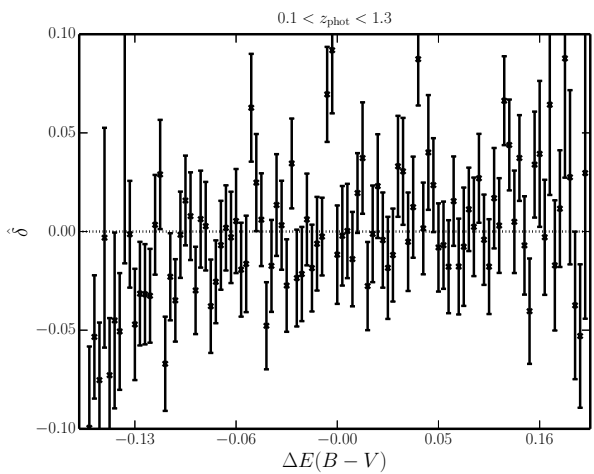}
    \caption{\label{fig:z01t13_ext_den} Galaxy over-density
      $\hat\delta$ against the relative $E(B-V)$ value from the SFD
      dust maps for the sample $0.1<z_{\rm phot}<1.3$.}
\end{figure}

\begin{figure*}
    \includegraphics[width=0.95\textwidth]{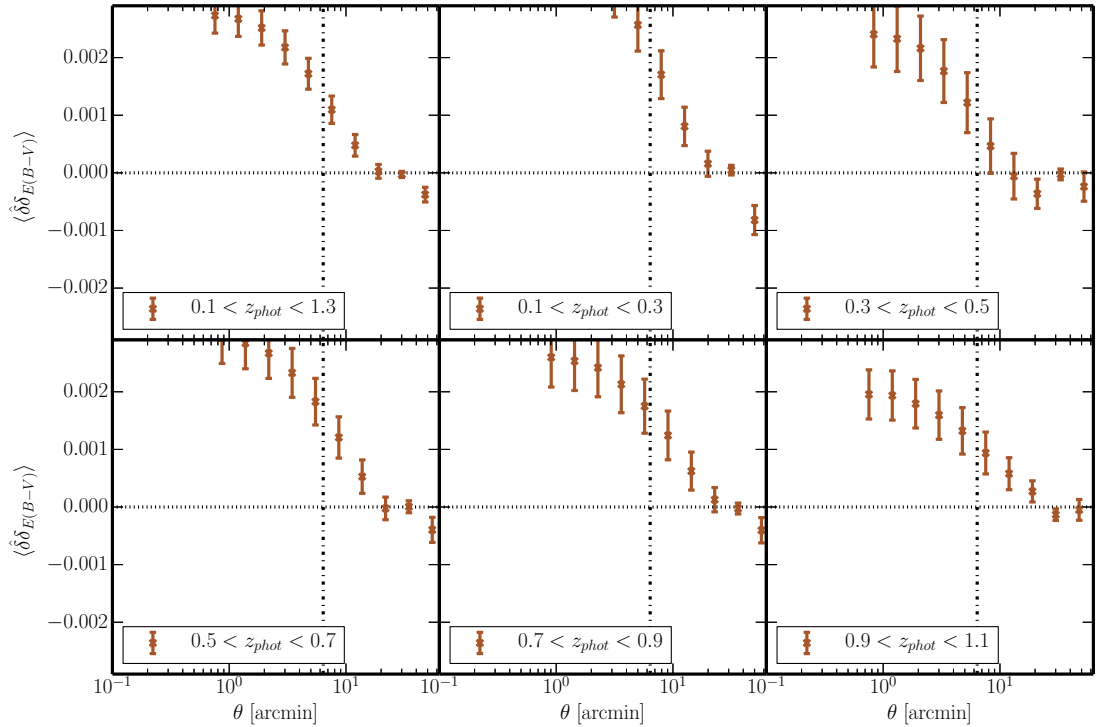}
    \caption{\label{fig:fore_ext}This figure is similar to
      Fig.~\ref{fig:fore_sky} except we now show the correlations of
      galaxy density at different redshifts against the galactic
      extinction $E(B-V)$ from the SFD dust maps \citep{schlegel98}.}
\end{figure*}

Fig.~\ref{fig:z01t13_ext_den} shows the density of the $0.1<z_{\rm
  phot}<1.3$ sample plotted against the amount of extinction relative
to the mean of the pointing. The density variation shows a trend from
small values to large with the density increasing with extinction. The
reason is probably again Eddington bias as discussed in
Sect.~\ref{sec:i_band_selected}. After correction for the reddening
the main remaining effect of dust is that it changes the depth of the
data. This systematic mostly affects the lower redshifts as their
bluer colours make them more sensitive to dust absorption.

The cross-correlations of the galaxy densities in different redshift
slices against the dust map are shown in
Fig.~\ref{fig:fore_ext}. Large amplitudes are found suggesting that
the effect of dust changing the noise properties of galaxies is at
least as important as the effects from varying depth and
seeing. Surprisingly, all samples show a significant anti-correlation
at large scales ($\sim50'$) on the order of $\sim0.05\%$.

\begin{figure}
    \includegraphics[width=0.495\textwidth]{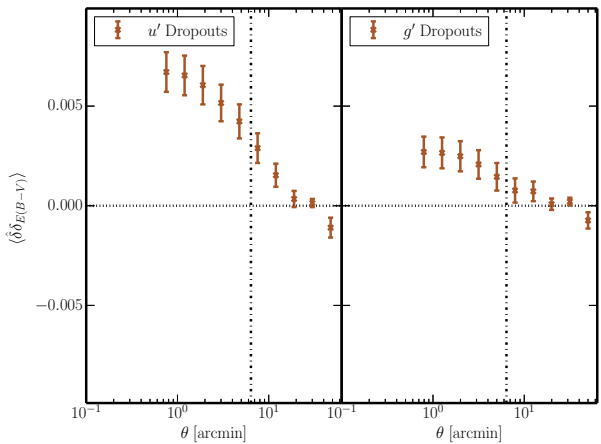}
    \caption{\label{fig:back_ext}This figure is similar to
      Fig.~\ref{fig:fore_ext} except we show the correlation for the
      $u$- and $g$-dropout samples.}
\end{figure}

In Fig.~\ref{fig:back_ext} we show the correlations of the densities
of the LBG samples against the extinction maps. Unlike for
cross-correlation against the depth maps, where the LBG samples showed
amplitudes larger by an order of magnitude compared to the photo-$z$
selected samples, we find that the amplitude of the cross-correlation
of the extinction with the LBG densities is within a factor of 2-3
compared to the amplitude for the photo-$z$ selected samples. While
the LBG samples have the reddest colours of any sample we consider,
making them less sensitive to extinction, they rely on non-detections
in the bluer filters for their selection making them sensitive to the
extinction.

\subsection{Galaxy Density vs Stellar Density}\label{sec:stellar_den}

The density of stars can also impact the selection function of galaxy
samples \citep[see e.g.][]{ross12}. Two effects that can cause this
are problems with star-galaxy separation and stellar light halos
obscuring galaxies and changing their colours. Poor star-galaxy
separation causes stars to be miss-identified as galaxies and vice
versa. The effect is to contaminate the galaxy correlation signal with
the stellar correlation and suppress the amplitude. Galaxies observed
through stellar halos in dense stellar fields also have their
densities and correlations affected.

We attempt to quantify the amount of correlation between a stellar
sample, galaxies in CFHTLenS, and the other systematics, however we
find very inconsistent results based on how the stellar sample was
selected either through colour \citep{hildebrandt12} or purely
shape-based methods \citep{miller13}. These different stellar
catalogues correlate very differently against a given galaxy sample
for all magnitudes of the stellar sample.

We employ an alternative method to test the effect of stellar halos by
adding larger masks around each of stars from both stellar
samples. The masks are chosen to reject areas 100 times the area of
detected stellar objects. Little change is observed in the amplitude
of the cross-correlations at all scales. The higher redshift samples
show no change with the new more aggressive stellar masking. This and
the observed flatness of the angular correlation function suggest that
the cross-correlation of the stellar samples and the galaxy samples is
dominated by stellar contamination in the galaxy sample rather than
the previously mentioned halo effect. The stellar contamination is
part of a larger issue of star-galaxy separation that we do not
address in this paper.

These inconclusive results are the reason why we do not consider
stellar density as an additional systematic effect in our analysis in
the following. More generally, any systematic that can not be mapped
with sufficient accuracy is problematic and can not be
analysed/removed with our approach. The whole procedure relies on
detailed 2-dimensional maps of the systematics which are not readily
available for the stellar density.


\section{Modeling and Removing Angular Systematics}\label{sec:remove_sys}

Angular systematics from position-dependent selection functions can be
a significant portion of the observed signal when performing a
correlation analysis. In order to remove such spurious signals, we
employ an empirical method of creating weighted random catalogues from
the data.

Spectroscopic surveys already employ a weighting of galaxies to
correct for changes in sensitivity and selection function
\citep[e.g.][]{newman13, lefvre05, ross12}. These weightings are both
analytically and empirically determined and enable the use of random
galaxy samples that, when used in correlation estimators, remove much
of the systematic variation from the dataset. This is not as commonly
done with photometric surveys as selections are thought to be more
homogeneous. As we have shown above this is not the always the case.

Simulating the selection effects would be one way to account for
this. However, selections that use photometric redshifts estimated
from non-linear fitting are quite complex and difficult to
simulate. We can, however, remove such systematics from photometric
samples using a fully empirical method. Empirical modeling of
systematic effects in photometric correlations has been used in
Fourier-space for large-scale BAO analyses
\citep{ho12,leistedt2013,leistedt2014}. Here we describe a method that
works in configuration space and at small scales ($\sim
1$arc-minute). We estimate weight maps for random points from the data
itself using a mapping from measured instrumental/astrophysical
quantities to observed galaxy density. Here we consider the effects of
depth in five bands, PSF size in five bands, and galactic extinction
totaling in 11 survey systematics for CFHTLenS. This method can only
work if the quantities used are not intrinsically correlated with the
observed samples, i.e. through a process that we would like to
measure. As long as we limit ourselves to instrumental and galactic
quantities, extragalactic measurements should not be affected.


\subsection{Weighted Random Points}

Weighted random points can be used in order to remove position
dependent systematics induced by a varying selection function. This
approach attempts to model and remove the systematics by mimicking the
selection function's effect through random points that follow the
under- and over-densities. These random points are used in the
\citet{landy93} correlation estimator which can be written as
\begin{equation}
 \hat{w}(\theta) = \frac{D_1D_2 - D_1R_2 - R_1D_2 + R_1R_2}{R_1R_2}
\end{equation}
where $D_1D_2$ are the number of pairs as a function of angle between
the two data samples in the cross-correlation, $D_1R_2$ and $D_2R_1$
are the number of pairs between the data and samples of random points,
and $R_1R_2$ are pairs between the two random samples. If there is no
position dependent selection function the random catalogues are only
required to follow the same geometry as the data catalogues. It should
be noted that for auto-correlations $1=2$ but we generalize for
cross-correlations. This and similar estimators are designed to
measure excess densities relative to a random distribution of
points. By having the random samples mimic the density variations due
to the selection function, we can remove these contributions from the
correlation. In the above equation $R_1$ would ideally have the same
selection efficiency as $D_1$ and likewise for $R_2$ and $D_2$. {It
is advantageous to weigh the random samples instead of the data samples
because this approach decreases shot noise and maps the full selection
function rather than just at points sampled by the galaxies.}

These random points are difficult to construct both analytically and
empirically as they require a mapping of the values of the systematics
to the expected under and over-densities. Analytically determining the
mapping of systematics to the induced selection requires an
understanding of how the systematics affect the selection of the
galaxy sample considered. This is only possible for the simplest of
selections and becomes impossible without large assumptions when the
selection criteria are complex and the systematics numerous. In
principle one could attempt to simulate the data and selection process
that results in the systematics variations. However, this would
require complex end-to-end simulations of the full galaxy sample,
reduction- and detection-pipeline, multi-colour photometry, photo-z,
etc.

Determining the weighting empirically (or blindly) through mapping
galaxy density to the set of systematics values is a more realistic
option for complex selections. We can measure the value of each
systematic at every position in the survey and map it to the observed
density. This is the approach we take and it works well as long as
there is a sufficient density of galaxies to map out the
dependency. Once the galaxy sample drops below $\sim 1$ gal/arc-minute
in CFHTLenS the fidelity of the map suffers.  Another difficulty is if
our knowledge of the systematic does not have a fine enough resolution
to probe the scales of interest or is contaminated (as in the case of
the stellar density discussed above). In this paper, we find that with
the majority of galaxy samples we are able to fit for much of the
systematic variation.


\subsection{Data Preparation}\label{sec:tech-prep}

Creating a weighted masking for a photometric dataset such as CFHTLenS
is complicated and requires careful consideration. This is mainly due
to the large volume of data and the numerous possible systematics
making determining the weighting challenging if the galaxy sample
under consideration is sparse and the effect is small. The problem is
to estimate the density of galaxies as a function of 11 variables
making this a very high-dimensional problem.

Throughout this analysis we utilize the spherical pixelisation code
STOMP and follow a methodology similar to \citet{scranton02} to create
the maps of systematics as a function of position and compute the
correlations. We start by describing how the data is prepared, moving
on to the smoothing of the data and then the creation of the weight
maps.

We utilize the STOMP maps created and described previously in
Sect.~\ref{sec:data} to map the positional dependence of both the
survey systematics and galaxy densities. Specific to our analysis of
CFHTLenS, we use relative values of the density and systematics
compared to the mean of each pointing instead of the global mean of
the survey. This increases the signal to noise of our estimates in
CFHTLenS (see Fig.~\ref{fig:den_compare}). In order to properly
estimate the density of galaxies against the systematics, we must find
a way to bin the data such that the bins capture much of the variation
in the systematics but contain enough galaxies to properly estimate
their density. This is a classical problem in high-dimensional data
analysis and requires an algorithm to efficiently map the parameter
space and group data points. To do this we utilize a k-means
clustering approach to identify portions of the data that have similar
systematics values across pointings and estimate the galaxy density in
these clusters.


\subsection{k-means Clustering}

Binning high-dimensional data efficiently is a non-trivial problem. We use a 
k-means clustering method \citep{macqueen1967} in order to bin survey pixels 
with similar systematic values and compute an average of the sample galaxy 
density contained in those pixels. In general, this machine learning algorithm 
aims to locate portions of the data with equal variance and bin them into 
clusters. The algorithm attempts to minimize
\begin{equation}
\sum_{i=1}^{k}\sum_{x_j \in S_i} \| x_j - \mu_i \|
\end{equation}
where $\mu_i$ is the $i$th cluster mean, the point $x_j$ is a member
of the set $S_i$ of all points nearest to $ \mu_i$, the first sum is
carried out over all $k$ clusters, and the second sum is carried out
over all points in the $i$th cluster. Cluster centers are initially
chosen at random. The center is then updated using the mean of the
nearest data points. The process is repeated until the difference in
the sum above for new centers from one iteration to the next reaches a
convergence criterion. We pick this method due to the ease of
attributing new values to the already determined clusters and the
simplicity of interpolating as a function of distance from cluster
centers. We use the MiniBatchKMeans method from the scikit-learn
Python package\footnote{http://scikit-learn.org/} designed for fast
convergence on large datasets with a high dimensionality.

As this technique is minimizing the variance of the clusters, we
normalize the variance of each systematic considered to unity, that
is, each value of the systematics is normalized by
\begin{equation}
  z(\vec{\theta}) = \frac{s(\vec{\theta}) - \langle s_i
\rangle}{\sigma(s(\vec{\theta}) - \langle s_i \rangle)}
\end{equation}
where $z$ is the normalized value of the systematic at position
$\vec{\theta}$, $s$ is the original value of the systematic, $\langle
s_i \rangle$ is the average value of the systematic in pointing $i$
containing $ \vec{\theta}$, and $\sigma(s(\vec{\theta}) - \langle s_i
\rangle)$ is the standard deviation of the systematic minus the
pointing mean over the full survey. The k-means algorithm then
considers each dimension equally rather than over fitting certain
variables due to their small variance. This allows the method to fit
the systematics without any assumption about the galaxy sample's
sensitivity to a given systematic. On application to the CFHTLenS
dataset the number of clusters we use varies with the density of the
galaxy sample. We use three different numbers of clusters: 64, 128,
256, which are arbitrarily chosen but span the range of under and over
fitting the variations in data when removing the systematics. We then
calculate the average over-density and errors in each k-means cluster
by spatially jackknifing over each of the $\le$171 pointings of
CFHTLenS contributing to the cluster.


\subsection{Testing}\label{sec:tech-tests}

We use the mapping of systematics to densities that comes out of the
k-means clustering approach as a weighting from which to draw random
samples. We can test the success of our weighting scheme by comparing
the cross-correlations of the galaxy samples under consideration
against each of the different systematics maps both before and after
the weighting is applied. If the correction has worked, the
correlation between the galaxy samples and the systematics maps should
equal the correlation between the weight maps and the systematic maps
within shot noise errors. This is a result independent of the science
goal. The difference between these two correlations yields an unbiased
estimate of the amount of residual systematic error in the measurement
after the weighting has been applied. It also gives an objective
criterion to discard samples that can not be properly corrected using
this method as it shows they have more issues with possibly unknown
systematics than can currently be accounted for. These tests are also
independent of cosmology or any assumptions pertaining to the galaxy
sample as long as just instrumental and galactic systematics are
considered.


\section{Resulting corrections}\label{sec:results_cor}

Here we present results from modeling the relation of galaxy density
against the 11 different systematics determined from our k-means
method. {We show how well the weights we create remove the systematic
compared to the amount of systematic correlation we see before and after the
correction is applied. We do this in a manner similar to \citet{ho12} where the
expected correlation amount for our cross-correlation is
\begin{equation}\label{eq:corr_ratio}
\langle \delta_1 \delta_2 \rangle_{sys_{ij}} \sim \frac{\langle \delta_1
\delta_{sys_i} \rangle \langle \delta_2 \delta_{sys_j} \rangle}{\langle
\delta_{sys_i} \delta_{sys_j} \rangle}
\end{equation}
where $\langle \delta_1 \delta_2 \rangle_{sys_{ij}}$ is the contribution
from systematics $i$, $j$ contributing to the galaxy cross-correlation,
$\delta_{1/2}$ is the over-density of galaxies for samples 1 and 2
respectively, and $\delta_{sys_i}$ is the $i$th systematic considered. This
equation is only truly valid for linear correlations and is likely not
completely valid for the scales we consider. However, it does give us some
indication of success. This is the best we can do currently without observing
the galaxy cross-correlation before and after correction. This could add
confirmation bias as we may stop once this galaxy cross-correlation looks
''correct''. The plots shown in this section are only for the cases where
$i=j$. The cross-terms are left out of the paper for brevity but are observed
to be as improved as the diagonal terms.}

As an application of this method we compare the residual systematic correlation
signal to the expected signal from weak lensing magnification bias. We determine
that we have successfully mitigated the systematic if the majority of the
corrected correlations are consistent to within 10\% of the expected
magnification signal and that we see a reduction in amplitude for each
correlation. It is clear that such a criterion does not directly guarantee that
the correlation functions used in the science analysis are free from any
systematics. However, it is a necessary criterion to assure that none
of the systematics investigated here swamp the signal.

\begin{figure*}
    \includegraphics[width=0.995\textwidth]{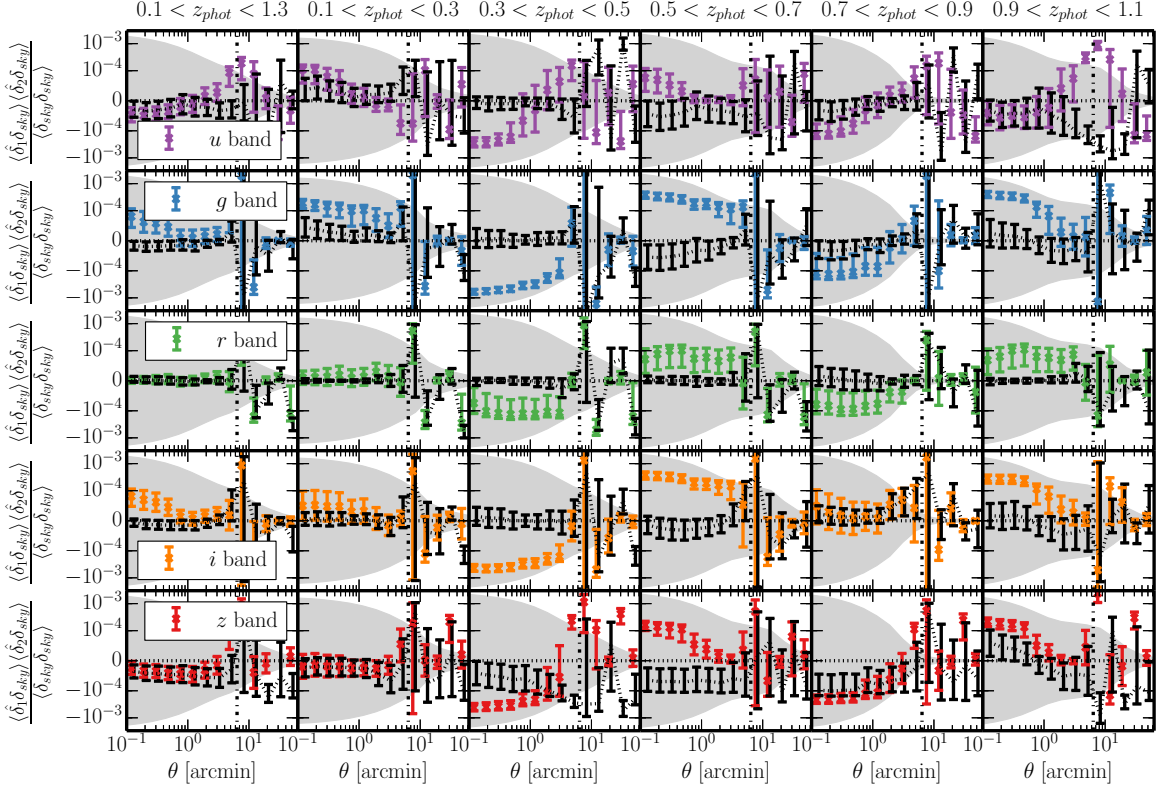}
    \caption{\label{fig:fore_sky_fix}{Symmetric log plots of the
observed and      corrected depth systematics for foreground lens galaxies
      cross-correlated against the $u'$ dropout sources. The $y$ axis is
      plotted linearly for the range $\pm10^{-4}$. The quantity plotted on     
      is the estimated contribution  from the depth to the magnification
      cross-correlation assuming linear systematics only. Unconnected $x$'s 
      show the raw systematic contamination. The black dashed line is the 
      estimated correlation after the correction is applied again assuming 
      linear systematics. Columns show these correlations for the different 
      photo-$z$ selected samples and the rows show the correlations against the
      different bands. The dark shaded region is $\pm$10\% of the
      expected magnification signal for the photo-$z$ sample (assuming
      a galaxy bias of $b=1$) lensing the $u$-dropout sample (assuming
      a slope of the number counts of $\langle \alpha - 1 \rangle=1$).}}
\end{figure*}

Fig.~\ref{fig:fore_sky_fix} shows the cross-correlation amplitude between the
lens galaxy samples and the $u$ Dropouts sample caused by the systematics
(unconnected coloured $x$'s) and this same systematic correlation after the
weighted correction is applied. This difference is shown
by the black data points connected by the dotted line. The dark shaded
region in each plot represents our criterion of 10\% of the
magnification bias signal expected from a cross-correlation of the
photo-$z$ sample and a source sample at $z=3$. For this signal we also
have to assume some value for the slope of the source number counts
and here we choose this to be $\langle \alpha - 1
\rangle=1$.\footnote{This is a conservative value for both dropout
  samples which are measured to have a slightly higher value
  \citep{hildebrandt09b}.} Additionally we assume a galaxy bias of
$b=1$ for the foreground sample.

\begin{figure*}
    \includegraphics[width=0.995\textwidth]{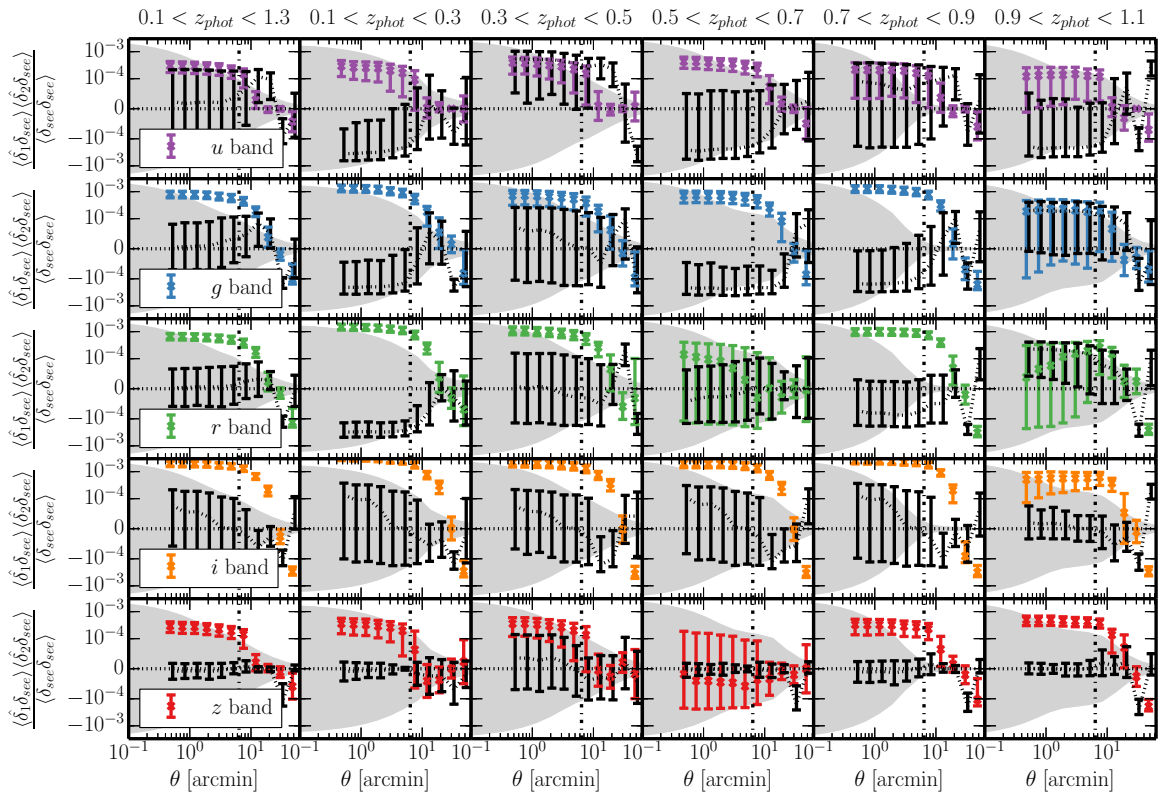}
    \caption{\label{fig:fore_see_fix}Same as
      Fig.~\ref{fig:fore_sky_fix} but showing the correlations against
      the PSF FWHM maps instead of the survey depth.}
\end{figure*}

\begin{figure*}
    \includegraphics[width=0.995\textwidth]{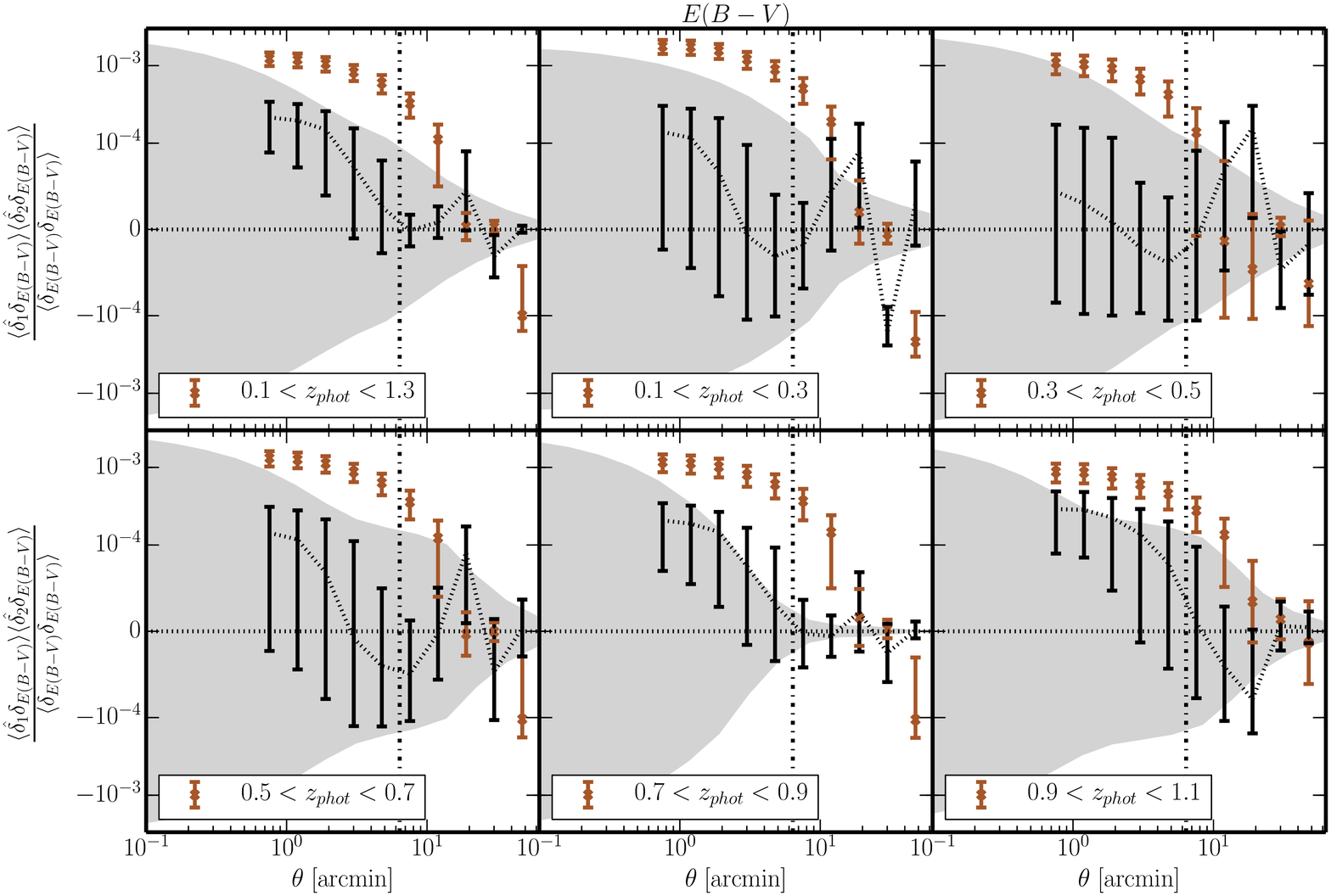}
    \caption{\label{fig:fore_ext_fix}Similar to
      Fig.~\ref{fig:fore_sky_fix} but for correlations against the
      $E(B-V)$ map as derived from the SFD dust
      maps\citep{schlegel98}. For these plots, each separate panel is
      for a different galaxy sample.}
\end{figure*}

{
For the contribution from the depth, the photo-$z$ selected samples with
redshifts $z_{\rm phot}<0.9$ and the global redshift sample ($0.1<z_{\rm
phot}<1.3$) correlated against the $u$-dropout sample are mostly
consistent with the 10\% criterion. For higher redshifts this is not the
case, especially in the $u$-band. The $0.3<z_{\rm phot}<0.5$ and $0.5<z_{\rm
phot}<0.7$ correlations show the most marked improvement, where the amplitudes
before correction are the largest of all of photo-$z$ samples and the amplitude
after correction is reduced to a level consistent with zero at small scales
($\theta<5$ arc-minutes) for all but the $z$-band. In general, all the
correlations are consistent with zero after correction at scales $\theta<10$
arc-minutes. At large scales it is difficult to interpret the results due to
the noise in these bins at the edge of the survey pointings.

For the higher redshift slices, the improvement is not
as dramatic. The $0.9<z_{\rm phot}<1.1$ sample shows that correlations
vs. the survey depth are improved in all bands except for the
$u$-band which retains a large coherent amplitude at $\sim10$
arc-mintues. Explaining this discrepancy is difficult considering the marked
improvement in the other four bands. It is likely due to this sample being
significantly fainter and sparser than the other samples considered. This both
makes the systematics more severe due to the faintness and their effect harder
to estimate due to the sparseness of the sample.

The improvements seen for the low $z_{\rm phot}$ cross-correlations comes from
both the foreground lens and source $u$-dropout LBGs. The LBGs show the
largest correlations for all the systematics, especially the $u$ dropout
sample. These source samples show improvements by an order of magnitude from
what is observed in Figs.~\ref{fig:back_sky},~\ref{fig:back_see},
and~~\ref{fig:back_ext} and are a large part of the improvement seen when
combined with the foreground. The results shown here are similar for the
$g$-dropouts which are left out of this section for brevity.

Fig.~\ref{fig:fore_see_fix} shows the correlations and corrections against the
seeing for the photo-$z$ selected and LBG samples. We utilise the
same empirically estimated density map as in the previous figures which is
simultaneously correcting for all the considered systematics. The systematics
observed here are roughly on the same order as the depth systematics with more
of the seeing systematic correlation amplitudes being outside of our criteria.
After correction, the correlation amplitudes show marked improvement with the
$z$-band and $i$-band being consistent with zero. The other bands are slightly
less successful but are made consistent with our criteria.

Fig.~\ref{fig:fore_ext_fix} shows the extinction map correlated against the
different CFHTLenS galaxy samples and the corrected correlations. Modelling of
this systematic is the least successful, though for the lower redshift ($z_{\rm
phot}<0.9$) galaxies the amplitude is reduced by an order of magnitude for most
samples. One possible reason for this is that the extinction is the only
variable considered that is contaminated by signal from large-scale
structure. The $E(B-V)$ extinction maps are in part determined by
infrared emission and partly contaminated by emission from the cosmic
infrared background as shown in e.g. \citet{schmidt15}.

Overall, we find encouraging results in modeling and removing the
largest systematics, reducing much of the spurious signal from the 11
systematics considered to consistent with zero for the expected contribution to
the signal from systematics. This is largely true for the lower redshift bins,
i.e.  samples with $z_{\rm phot}<0.9$. For higher photo-$z$ slices, the
corrections and modeling tend to break down. This is likely due to the
increased sparseness and faintness of these samples with the mean magnitude
shifting by 1 magnitude in the most extreme case. These samples also
have less that 1/3rd of the galaxies as the number of lower redshift samples.
These faint samples are much more susceptible to the effect of systematics
and it is likely there is not enough data to constrain these effects.
We find more success with the Lyman-break galaxy samples likely due to
the simple nature of the colour-colour selection. Large-scale
corrections can be problematic for some samples though this is difficult to
interpret due to noise. This is likely due to
smoothing we perform, treating each pointing individually rather than
the whole survey simultaneously.}


\section{Discussion and Conclusions}\label{sec:conclusions}

{
In this paper we have presented an investigation into systematic
density variations of galaxy samples in CFHTLenS caused by variations
of instrumental and astrophysical quantities. We also present a model
that attempts to mitigate these effects, empirically modeling galaxy
density as a function of survey systematics. We model these survey
systematics with an eye toward measuring magnification in CFHTLenS and
other similar multi-epoch, wide area surveys. We consider survey
systematics of limiting magnitude (depth), PSF size (FWHM), and
galactic extinction ($E(B-V)$). In total for CFHTLenS this is 11
systematic variables: 5 depth, 5 PSF sizes, and 1 extinction.  We find
that, without a correction, correlations induced by variations in such
systematics can be significant compared to the intended
measurement. We find induced density fluctuations in small fractions
of the area of the survey of up to 10\% in photo-$z$ selected samples
and up to even 50\% in high-$z$ Lyman-break galaxy samples. This can
result in correlation amplitudes of 0.1\% and 1.0\%, respectively. In
the case of magnification bias, this spurious correlation can be as
large as twice the expected amplitude of the signal one wants to
measure!

We modeled these systematics by identifying regions of similar
systematic values in the survey (in the 11 dimensions mentioned above)
using a k-means clustering approach and estimating the average galaxy
over-density as a function of the survey systematics. Using this
method we see improvement in the spurious cross-correlation
systematics for the majority of the galaxy samples considered. In some
cases, the weighted correction reduces the expected contribution from
systematics such that it is consistent with zero. The high redshifts,
unfortunately, do not improve considerably and still show significant
correlation of similar amplitude to the magnification. This is likely due to the
large difference in brightness of the samples, with the higher redshift samples
being much fainter and sparser. These fainter samples are then more sensitive to
selection effects and, due to these samples being sparser and noisier,
we are unable to effectively estimate a correction with high
precision.

The LBG samples considered are an exception. While they are the
faintest and sparsest samples that we are trying to correct, their
corrections work surprisingly well. This is likely due to the fact
that the LBGs are selected with simple, two-dimensional colour
cuts. The high-$z$ photo-$z$ samples in contrast are selected in a
much more complicated way by the inner workings of a photo-$z$ code
($BPZ$ in our case) so that it is conceivable that their dependence on
systematics is also more complicated.}

The analysis presented here directly enables measurements of
magnification bias in the CFHTLenS. In the future, we will apply this
technique to other surveys such as RCSLenS and KiDS. These surveys are
very similar to CFHTLenS (especially in terms of data processing) and
will enable quick turn around on the science.

This technique will benefit from larger area, deeper future surveys,
however, care will need to be taken to assure that computation time is
tractable. The k-means algorithm used in this work is designed for
large datasets and will likely scale properly though further
investigation is needed. In order to proceed with these larger
datasets and properly remove systematics down to smaller scales,
simulations of the dataset mimicking the observing strategy and
large-scale structure will be required. This will give an
understanding of both density dependence on survey systematics, but
also systematics from changes in densities caused by deblending and
light halos. {This method could then be trained with simulated
catalogs that mimic the systematics probed. This is left for future work.}

The fundamental limitation of this technique is the finite size of the
galaxy sample that is used to map the dependency on the different
systematics. While larger surveys will contain more information to
establish this relation those surveys will also require systematic
errors to be controlled to a higher level.  It is not clear at the
moment which of the two aspects will win in the end, i.e. whether our
ability of modelling systematic effects will scale more favourably
with increasing data volume than statistical noise or the opposite.

\section*{Acknowledgements}

We thank Peter Schneider for his valuable feedback. We also thank our referee,
Enrique Gaztanaga for his helpful comments and suggested improvements of the
manuscript. CM and HH are
supported by the DFG Emmy Noether grant Hi 1495/2-1.

This work is based on observations obtained with MegaPrime/MegaCam, a joint
project of CFHT and CEA/IRFU, at the Canada-France-Hawaii Telescope (CFHT) which
is operated by the National Research Council (NRC) of Canada, the Institut
National des Sciences de l'Univers of the Centre National de la Recherche
Scientifique (CNRS) of France, and the University of Hawaii. This research used
the facilities of the Canadian Astronomy Data Centre operated by the National
Research Council of Canada with the support of the Canadian Space Agency.
CFHTLenS data processing was made possible thanks to significant computing
support from the NSERC Research Tools and Instruments grant program.

\bibliographystyle{mn2e}
\bibliography{cfht_sys}

\end{document}